\documentclass[journal]{IEEEtran}
\usepackage{lineno}
\usepackage{cite}
\usepackage{hyperref}
\usepackage{amsmath,amssymb,amsfonts}
\usepackage{amsmath}
\usepackage{amsthm}

\usepackage{graphicx}
\usepackage{textcomp}
\usepackage{xcolor}
\usepackage{graphicx}
\usepackage{float}
\usepackage{subfigure}
\usepackage{amsmath}
\usepackage{amsfonts,amssymb}
\usepackage{mathrsfs}
\usepackage{mathtools}
\usepackage{multirow}
\usepackage{algpseudocode}
\makeatletter
\newif\if@restonecol
\makeatother

\usepackage[linesnumbered,ruled,vlined]{algorithm2e}
\usepackage{algpseudocode}

\usepackage{setspace}
\usepackage{footmisc}
\usepackage[justification=centering]{caption}

\usepackage{mathtools}
\usepackage{dsfont}
\usepackage{bbm}

\newtheorem{corollary}{Corollary}

\theoremstyle{definition}
\newtheorem{theorem}{Theorem}

\newtheorem{lemma}{Lemma}

\makeatletter
\newcommand{\biggg}{\bBigg@{3}}
\newcommand{\Biggg}{\bBigg@{3.5}}
\makeatother
\usepackage{bm}
\makeatother
\begin{document}

\title{Uplink Sum-Rate Maximization for Pinching Antenna-Assisted Multiuser MISO Communications}
\author{
Jiarui Zhang, Hao Xu, Chongjun Ouyang, Qiuyun Zou, Hongwen Yang
}

\maketitle

\begin{abstract}
This article investigates the application of pinching-antenna systems (PASS) in multiuser multiple-input single-output (MISO) communications. Two sum-rate maximization problems are formulated under minimum mean square error (MMSE) decoding, with and without successive interference cancellation (SIC). To address the joint optimization of pinching antenna locations and user transmit powers, a fractional programming-based approach is proposed. Numerical results validate the effectiveness of the proposed method and show that PASS can significantly enhance uplink sum-rate performance compared to conventional fixed-antenna designs.
\end{abstract}

\begin{IEEEkeywords}
Multiuser communications, pinching antennas, sum-rate maximization, uplink transmission.
\end{IEEEkeywords}

\section{Introduction}
\IEEEPARstart{T}{he} advent of 6G communication systems calls for the application of flexible-antenna technologies, such as fluid antennas \cite{FluidAntennaSystems} and movable antennas \cite{movableantenna}. These technologies have some limitations that they cannot have a large-scale antenna reconfiguration and lack the flexibility to add or move antennas.

To solve these problems, pinching antennas (PAs) has introduced by NTT DOCOMO in 2021 \cite{pinchingantenna, Fukuda2022}, which leverage dielectric waveguides for flexible transmission as shown in Fig. \ref{fig_1}. Electromagnetic waves are emitted by pinching small dielectric particles at specific points along the waveguide \cite{ouyangArrayGainPinchingAntenna2025}. Similar to adding or removing a clothpin from a clothline, these dielectrics are typically attached to the tips of plastic clips and can be dynamically added or removed from the waveguide for precisely serving wireless communications\cite{liuPinchingAntennaSystemsPASS2025}.

Recent studies have highlighted several advantages of PAs\cite{yangPinchingAntennasPrinciples2025}. Unlike the conventional flexible-antenna systems, PAs mitigate large-scale path loss by adjusting PA placement across waveguide. This procedure referred to as "pinching beamforming", which enhance network coverage and ``last-meter'' connectivity \cite{liuPinchingAntennaSystemsPASS2025}. Besides, PAs offer scalability in adding or removing antennas and reconfigure the channel environment at low cost and low complexity. These take the benefits of outperforming traditional fixed-antenna designs in large coverage areas. In essence, PASS can be viewed as a specific implementation of fluid-antenna or movable-antenna concepts, which offers a more flexible and scalable solution than traditional architectures. In recognition of DOCOMO’s contribution, we refer to this technology as PASS throughout this paper.

These benefits have spurred significant research into pinching-antenna systems (PASS). Up to now, most studies considered downlink transmission. The work in \cite{dingFlexibleAntennaSystemsPinchingAntenna2024} was the first theoretical study on the PASS and focused on the practical designs. This research considered about a multiuser multiple-input single-output (MISO) PASS with one waveguide and multiple PAs, and the result showed that PASS can motivate the application of nonorthogonal multiple access. Meanwhile, the authors in \cite{xuRateMaximizationDownlink2025} optimized PA positioning for a downlink system, where multiple pinching antennas are deployed on a waveguide to serve a single-antenna user. The activation of PAs along one waveguide is further considered in \cite{wangAntennaActivationNOMA2024} to serve multiple users. To establish an accurate signal model, the authors in \cite{wang_modeling_2025} first indicated the electromagnetic field behavior and proposed a physics-based hardware model for PASS. Besides, some learning-based methods were proposed in \cite{guoGPASSDeepLearning2025}, \cite{xieGraphNeuralNetwork2025} to realize the efficient beamforming in PASS, which can be applied to the situation of one waveguide with multiple PAs.

The above studies validated the effectiveness of PASS in wireless communication. However, all these studies considered the downlink scenarios. For uplink transmission, the authors in \cite{houPerformanceUplinkPinching2025} and \cite{tegosMinimumDataRate2024} analysed the performance of uplink system with multiple users and orthogonal multiple access (OMA), respectively, which were limited to scenarios where all PAs are positioned within a single waveguide. It is noted that there is no research considering deploying multi-waveguides in uplink multiuser communications, which results in a joint optimization problem of users' power and pinching beamforming. Therefore, despite its importance, how to improve the uplink sum-rate for pinching antennas assisted multiuser systems with multiple waveguides still remains an open problem.

To fill the research gap, this paper investigates uplink transmission for multiuser multiple input single output (MISO) systems with PAs employed in multi-waveguides. Our contributions are summarized as follows: i) We propose the frame of uplink PASS with multiple waveguides. ii) We formulate a optimization problem respecting PAs' positions and users' power under two classic combining rules—minimum mean square error (MMSE) with successive interference cancellation (SIC) and without SIC\cite{tseFundamentalsWirelessCommunication2005}. iii) We propose an efficient fractional programming (FP)-based algorithm \cite{FP} in a structure of block coordinate descent (BCD) to jointly optimize PAs' positions and user power allocation. iv) We present numerical results to demonstrate that the proposed method outperforms conventional methods in sum-rate performance by increasing system flexibility.

\section{System Model And Problem Formulation }
\begin{figure}[!t]
\centering
\includegraphics[width=3.5in,trim=0in 0in 0in 0in, clip]{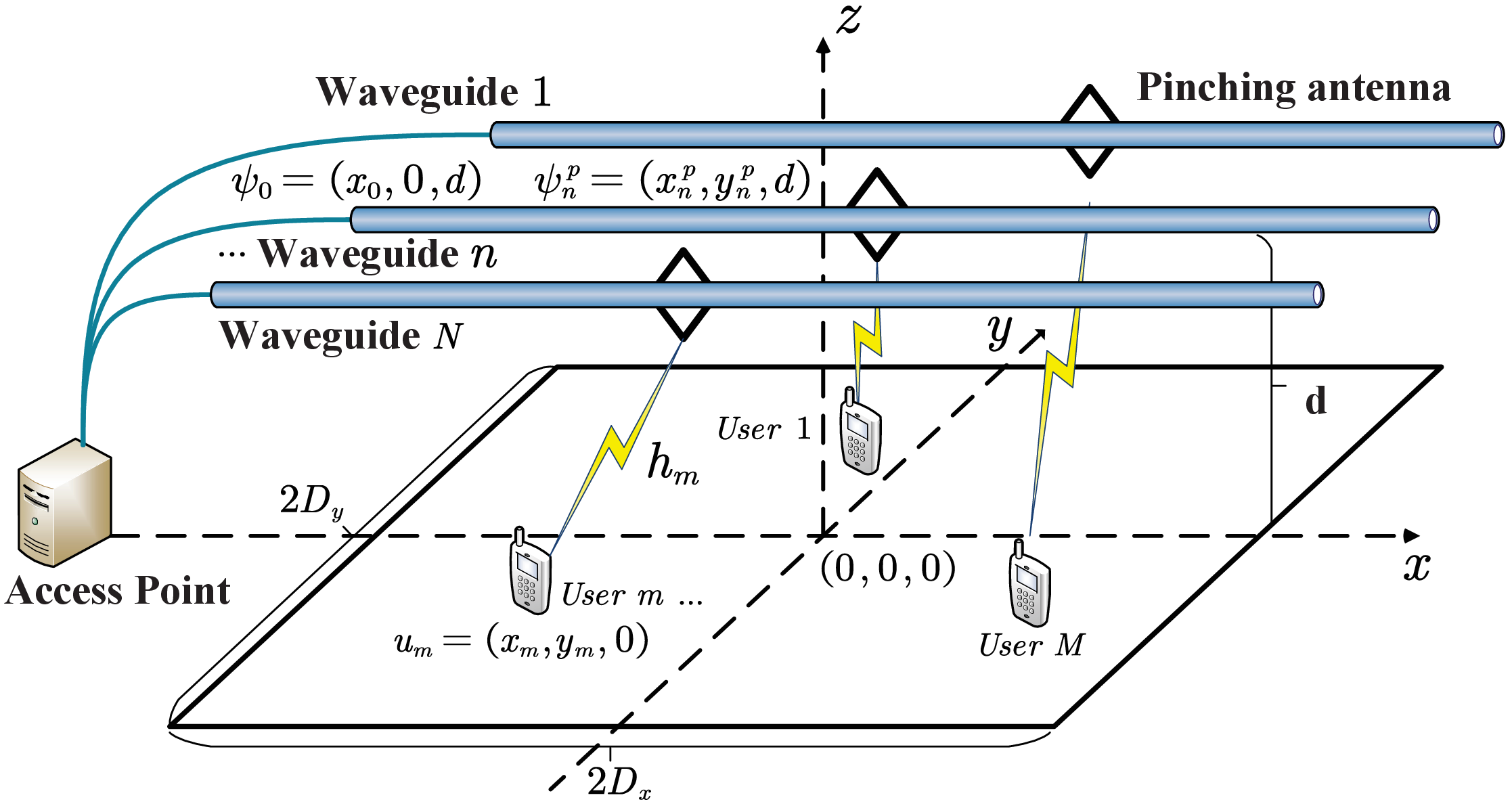}
\caption{Illustration of a PASS-assisted multiuser MISO uplink channel.}
\label{fig_1}
\end{figure}

Consider a PASS-assisted uplink MISO communication system consisting of $N$ dielectric waveguides and $M$ single-antenna users, as illustrated in Fig. \ref{fig_1}.

\subsection{Pinching-Antenna System}
We assume that each PA is activated on different waveguides, jointly serving the users. Each pinching element can freely move across different waveguides and transmit signal to the access point. Waveguides are extended over the $x$-axis at the altitude $d$ in an array formed on the $y$-axis with each two waveguides being distanced $\frac{D_y}{N}$. The location of pinching element $n$ on the waveguide $n\in[\mathcal{N}]$ in the Cartesian coordinate is given by $\bm{\psi}_{n}^p=[x_n^p,y_n^p,d]$, where $x_n^p\in[-D_x,D_x]$ and $y_n^p =-D_y+\frac{nD_y}{N}$.

The users are distributed within a known two-dimensional area, e.g., a rectangular region. The $m$th user is located at $\mathbf{u}_m=[x_m,y_m,0]^{\mathsf T}$ for $\forall m\in \mathcal{M}=\{1,\dots,M\}$ and transmits its signal to the access point located at ${\bm\psi}_0=[x_0,0,d]$ with $x_0< 0$. Specifically, $x_m$ and $y_m$ are subject to the constraints of $x_m\in[-D_x,D_x], y_m\in[-D_y,D_y]$, assuming that the users are located in the $xy$-plane.

The spatial channel between the $m$th user and the PASS can be expressed as follows \cite{dingFlexibleAntennaSystemsPinchingAntenna2024}:
\begin{align}
\mathbf{h}_m = [h_{m,1},\dots,h_{m,N}]^{\mathsf T} =
\bigg[\frac{\eta^\frac{1}{2}e^{-j\frac{2\pi}{\lambda}\|\mathbf{u}_m-\bm{\psi}_n^p\|}}{\|\mathbf{u}_m-\bm{\psi}_n^p\|}\bigg]^N_{n=1}
,
\end{align}
where $\eta=\frac{c^2}{16\pi^2f_c^2}$, $c$ denotes the speed of light, $f_c$ is the center frequency, and $\lambda$ is the wavelength in the free space equal to $\frac{c}{f_c}$. Considering about the phase shift of received signals by the $N$ PAs, we define $\bm{\phi}=[e^{-j\phi_1},\dots,e^{-j\phi_N}]^{\mathsf T}\in{\mathbb{C}^{N\times 1}}$, where $\phi_n=\frac{2\pi}{\lambda_g}(x_n^p-x_0)$, and $\lambda_g=\frac{\lambda}{n_{\textrm{eff}}}$, with $n_{\textrm{eff}}$ representing the effective refractive index of the dielectric waveguide\cite{DMMicrowave2021}.

Taken together, the effective channel between the $m$th user and the PASS can be expressed as $\mathbf{g}_m = \bm{\phi}\circ \mathbf{h}_m$, where $\circ$ denotes the Hadamard product. The received signal at the PASS is given by
 \begin{equation}
\mathbf{y} = \sum_{m=1}^{M} \sqrt{p_m}\mathbf{g_m}s_m+\mathbf{n}, \label{signal}
\end{equation}
where $s_m \sim \mathcal{CN}(0,1)$ denotes the symbol transmitted by the $m$th user with $\mathbb{E}\{s_i s_j^{*}\}=0, \forall i \neq j$, $\mathbf{n} \sim \mathcal{CN}(0, \sigma^2 \mathbf{I}_N)$ is additive white Gaussian noise with power $\sigma^2$, and $p_m$ presents the transmit power of the $m$th user, subject to $p_m\leq p_{\text{max}}$, where $p_{\text{max}}$ is the maximum power. After receiving the signal, BS employ two different combining rules, MMSE-SIC and MMSE-nSIC combiner\cite{tseFundamentalsWirelessCommunication2005}. The sum-rate with these are given by \eqref{eq:SIC} and \eqref{eq:nSIC}, as shown at the top of next page. Note that the decoding order of MMSE-SIC method has no influence of the sum-rate. Thus, to simplify the analysis, we assume that the decoding order of MMSE-SIC method is from the $1$st user to the $M$th user.

\begin{figure*}[!t] 
\begin{equation}
\mathcal{R}_{\text{sic}} = \sum_{m=1}^{M} \log \left( 1 + p_m\mathbf{g}_m^H \left( \sum_{i=m+1}^{M} \mathbf{g}_i \mathbf{g}_i^H {p}_i + \sigma^2 \mathbf{I}_L \right)^{-1} \mathbf{g}_m \right),  \label{eq:SIC}
\end{equation}

\begin{equation}
\mathcal{R}_{\text{nsic}} \hspace{5pt}  = \hspace{5pt} \sum_{m=1}^{M} \log \left( 1 + p_m\mathbf{g}_m^H \left( \sum_{i\neq m}^{M} \mathbf{g}_i \mathbf{g}_i^H {p}_i + \sigma^2 \mathbf{I}_L \right)^{-1} \mathbf{g}_m \right),  \label{eq:nSIC}
\end{equation}
\rule{\linewidth}{1pt}
\end{figure*}

\subsection{Problem Formulation}
We aim to maximize the sum-rate of the considered uplink PASS by jointly optimizing the power of users and positions of PAs. The optimization problem can be formulated as follows:
\begin{subequations}
\begin{align}
\mathcal{P}_\text{X}: \max_{\mathbf{P}, \bm{\Psi^p}} & \quad \mathcal{R}_\text{X}  \\
\text{s.t.} \quad & 0\leq{p}_m \leq P_{\max}, \quad \forall m \in \mathcal{M},  \label{limit:p} \\
\quad & x_n^p \in [-D,D] \quad \forall n \in \mathcal{N}, \label{limit:x}
\end{align}
\end{subequations}
where $\mathbf{p}=\big[p_m\big]_{m=1}^M,\mathbf{\Psi}^p = \big[\psi_{n}^p\big]_{n=1}^N$, $\text{X} \in \{\text{sic},\text{nsic}\}$.

\section{Proposed Method}
\subsection{MMSE-SIC case}
Solving the optimization problem \(\mathcal{P}_{\text{sic}}\) is highly challenging due to several points. First, the objective function of \(\mathcal{P}_{\text{sic}}\) is non-convex with respect to the variables \(\{\mathbf{P}, \mathbf{\Psi}^p\}\). In addition, minor changes in antenna positions can result in significant phase variations at high-frequency bands, which indicates that positioning needs an optimization algorithm with higher accuracy. To develop a computationally efficient algorithm, we exploit the fractional programming framework \cite{FP} and introduce two sets of auxiliary variables to transform \(\mathcal{P}_{\text{sic}}\) into a new problem with a more manageable objective function, while maintaining the equivalence of the original problem. The following two lemmas lay the groundwork for this transformation.
\vspace{-5pt}
\begin{lemma}\label{Lemma1}
Problem ${\mathcal{P}}_{\text{sic}}$ is equivalent to
\begin{subequations} \label{P_1}
\begin{align}
\mathcal P_1:\max_{{\mathbf{P}},{\bm{\Psi}^p}}~&{\mathcal{F}_1}\left({\bm{\alpha}},{\textbf P},{\bm{\Psi}^p}\right)\triangleq
\sum\nolimits_{m=1}^{M}\log\left(1\!+\!\alpha_k\right)\!-\!\sum\nolimits_{m=1}^{M}\alpha_m\nonumber\\
&+\sum\nolimits_{m=1}^{M}\left(1+\alpha_m\right){\mathcal{A}}_m
\\
{\text{s.t.}}~&\eqref{limit:p},\eqref{limit:x},\alpha_m>0,m=1,\cdots,M, \label{limit:P1}
\end{align}
\end{subequations}
\end{lemma}
where ${\mathcal{A}}_m\triangleq{p_m}{\textbf{g}}_m^{\mathsf H}\left(\sum\limits_{i=m}^M {{{\textbf{g}}_i}{\textbf{g}}_i^{\rm H}{p_i} }+{\sigma ^2}{{\textbf{I}}_N}\right)^{-1}
{\textbf{g}}_m$, \\
${\bm{\alpha}}=\left[\alpha_1,\cdots,\alpha_M\right]^{\rm T}$, where the optimal $\alpha_m$ is given by
\vspace{-5pt}
\begin{align}\label{Optimal_alpha}
\alpha_m^{\dagger}\triangleq{p_m}{\textbf{g}}_m^{\mathsf H}\left(\sum\nolimits_{i=m+1}^M {{{\textbf{g}}_i}{\textbf{g}}_i^{\mathsf H}{p_i} }+{\sigma ^2}{{\textbf{I}}_N}\right)^{-1}{\textbf{g}}_m.
\end{align}
\vspace{-5pt}
\begin{IEEEproof}
{Let $\left\{{\bm \alpha}^{\dagger},{\mathbf P}^{\dagger},{\bm{\Psi}^p}^{\dagger}\right\}$ denote the optimal solution of $\mathcal P_1$. Then, $\max\limits_{{\bm \alpha},{\mathbf P},\bm{\Psi}^p}{\mathcal F_1}=\max\limits_{{\mathbf
P},{\bm{\Psi}^p}}{\mathcal F_1}\left({\bm \alpha}^{\dagger},{\mathbf P},{\bm{\Psi}^p}\right)$ and $\left\{{\mathbf P}^{\dagger},{\bm{\Psi}^p}^{\dagger}\right\}=\mathop{\rm argmax}\limits_{{\mathbf P},{\bm{\Psi}^p}}{\mathcal F}\left({\bm \alpha}^{\dagger},{\textbf P},{\mathbf{\Psi}^p}\right)$. Given the analysis of \(\mathcal{F}_1\) across the entire complex plane and its concavity with respect to \(\bm{\alpha}\) for a fixed set \(\left\{{\mathbf P},{\bm{\Psi}^p}\right\}\), we derive its complex gradient and resolve the condition \(\frac{\partial \mathcal{F}_1}{\partial \alpha_m} = 0\) for every index \(m\). Thus, the optimized $\bm \alpha$ is obtained as $\alpha_m^{\dagger}=\frac{{\mathcal{A}}_m}{1-{\mathcal{A}}_m}$, which can be derived as shown in \eqref{Optimal_alpha} using the Woodbury formula. Inserting $\alpha_m^{\dagger}$ back to $\mathcal F_1$ yields ${\mathcal F_1}\left({\bm \alpha}^{\dagger},{\mathbf P},{\bm{\Psi}^p}\right)={\mathcal{R}}_{\text{sic}}$. Hence, we can get $\max\limits_{{\bm \alpha}^{\dagger},{\mathbf P},{\bm{\Psi}^p}}{\mathcal F_1}
=\max\limits_{{\mathbf P},{\bm{\Psi}^p}}{\mathcal{R}}_{\text{sic}}$ and $\mathop{\rm argmax}\limits_{{\mathbf P},{\bm{\Psi}^p}}{\mathcal F_1}\left({\bm \alpha}^{\dagger},{\mathbf P},{\bm{\Psi}^p}\right)
=\mathop{\rm argmax}\limits_{{\mathbf P},{\bm{\Psi}^p}}{\mathcal{R}}_{\text{sic}}$, which suggests that ${\mathcal{P}}_{\text{sic}}$ and $\mathcal P_1$ have the same optimal objective value as well as the same optimal solution to $\left\{{\textbf P},{\mathbf{\Psi^p}}\right\}$.
}
\end{IEEEproof}
With the help of Lemma \ref{Lemma1}, ${\mathcal{P}}_{\text{sic}}$ becomes more manageable. However, it still involves a matrix inversion term, making it a non-convex problem. To further address this, we employ the FP-based approach and introduce auxiliary variables ${\bm \beta}=\{{\bm \beta}_1,\cdots,{\bm \beta}_M\}$ to simplify ${\mathcal F_1}\left({\bm \alpha},{\mathbf P},{\bm{\Psi}^p}\right)$. The lemma \ref{Lemma2} is shown as follows.

\vspace{-5pt}
\begin{lemma}\label{Lemma2}
Problem ${\mathcal{P}}_1$ is equivalent to the following:
\begin{subequations} \label{P_2}
\begin{align}
\mathcal P_2:\max_{{\mathbf{P}},{\bm{\Psi}^p}}~&{\mathcal{F}_2}\left({\bm \alpha},{\bm \beta},{\mathbf P},{\bm{\Psi}^p}\right)\triangleq \sum_{m = 1}^M \log(1 + \alpha_m) \nonumber \\
& - \sum_{m = 1}^M \alpha_m + \sum_{m = 1}^M \left[2\sqrt{1 + \alpha_m}\Re(\bm{\beta}_m^{\mathsf H} \mathbf{g}_m \sqrt{p_m}) \right. \nonumber \\
& \left. - \bm{\beta}_m^{\mathsf H} \left( \sum_{i = m}^M \mathbf{g}_i \mathbf{g}_i^{\mathsf H} {p_i} + \sigma^2 \mathbf{I}_N \right) \bm{\beta}_m \right] \label{P_sic_obj}
\\
{\text{s.t.}}~&\eqref{limit:P1}, \bm{\beta}_m \in \mathbb{C}^{N \times 1}, m \in \mathcal{M}, \label{limit}
\end{align}
\end{subequations}
\end{lemma}
where each $\bm{\beta}_m$ has the following optimal solution
\vspace{-5pt}
\begin{align}\label{Optimal_Beta}
{\bm\beta}_m^{\dagger}=\sqrt{1+\alpha_m}\left(\!\sum\nolimits_{i=m}^M\! {{{\textbf{g}}_i}{\textbf{g}}_i^{\mathsf H}{p_i}} \!+\!{\sigma ^2}{{\textbf{I}}_N}\!\right)^{-1}\!
\sqrt{{p}_m} {{\textbf{g}}_m}.
\end{align}

\vspace{-5pt}
\begin{IEEEproof}
The proof is similar to that of Lemma \ref{Lemma1}. Note that when ${\bm \alpha},{\mathbf P},{\bm{\Psi}^p}$ are fixed, $\mathcal{F}_2$ is concave with respect to $\bm\beta$. Therefore, by taking its complex derivative and solving the condition $\frac{\partial {\mathcal{F}_2}}{\partial \beta_m}=0$, the optimal solution for $\beta_m$ is $\beta_m^{\dagger}$ in \eqref{Optimal_Beta}. Following a similar approach as in the proof of Lemma \ref{Lemma1}, Lemma \ref{Lemma2} can be established.
\end{IEEEproof}
Based on Lemmas \ref{Lemma1} and \ref{Lemma2}, it is worth noting that problem $\mathcal P_2$ has a more tractable objective function than $\mathcal P_\text{sic}$, which is in terms of concerning each variable with others fixed. To solve problem $\mathcal P_2$, we propose a block coordinate descent algorithm by sequentially updating ${\bm{\alpha}, \bm{\beta}, \mathbf{P}, \bm{\Psi}^p}$.

\subsubsection{Optimize \texorpdfstring{$\bm \alpha$ and $\bm \beta$}{2}}
We first optimize $\bm\alpha$ with the other variables fixed. Specifically, by checking its first-order optimality condition, the optimal $\bm\alpha$ can be obtained, i.e., $\alpha_m^\dagger$ in (\ref{Optimal_alpha}). Similarly, by fixing other variables and checking their first-order optimality condition, the optimal solution to $\bm\beta$ is given by $\bm\beta_m^\dagger$ in \eqref{Optimal_Beta}.\\

\subsubsection{Optimize \texorpdfstring{${\bm\Psi}^p$}{2}}
Then we investigate the optimization of the PAs position by fixing other variables. The subproblem for ${\bm\Psi}^p$ is formulated as:
\begin{subequations}
\begin{align}
\mathcal P_3:\max_{{{\bm\Psi}^p}} & \quad {f_1}\left({{\bm\Psi}^p}\right)\triangleq \sum_{m = 1}^M \left[2\sqrt{1 + \alpha_m}\Re(\bm{\beta}_m^{\mathsf H} \mathbf{g}_m \sqrt{p_m}) \right. \nonumber \\
& \left. - \bm{\beta}_m^{\mathsf H} \sum_{i = m}^M \mathbf{g}_i \mathbf{g}_i^{\mathsf H} {p_i} \bm{\beta}_m \right] \label{P_psi_p} \\
{\text{s.t.}}~&\eqref{limit:x}. \label{limit:P3}
\end{align}

\end{subequations}

For each PA, the coordinate ${\bm\Psi}^p_n$ depends on ${x_n^p}$, and ${f_1}$ is a one-dimensional function for each ${x_n}$ when other PA positions are fixed. Specifically, \eqref{P_psi_p} can be reformulated as:
\begin{align}
~&{f_2}\left({{\bm\Psi}^p}\right)\triangleq \sum_{m = 1}^M \left[2\eta \sqrt{1 + \alpha_m}\sqrt{p_m}\Re(\sum_{n = 1}^N \overline{\beta_{m,n}}C_{m,n}) \right. \nonumber \\
& \left. - \sum_{i = m}^M \sum_{n = 1}^N  \sum_{n^\prime = 1}^N \eta p_i\overline{\beta_{m,n}}\beta_{m,n^\prime}C_{m,n}\overline{C_{m,n^\prime}}  \right], \label{P_xn_p}
\end{align}
where $C_{m,n}=\frac{e^{-j\phi_n}e^{-j\frac{2\pi}{\lambda}\|\mathbf{u}_m-\bm{\psi}_n^p\|}}{\|\mathbf{u}_m-\bm{\psi}_n^p\|}$ and $\beta_{m,n}$ denotes the $n$-th element of $\bm\beta_m$. Given the non-convexity of $f_2$, we adopt a the gradient decent method with backtracking
line search \cite{ConvexOp} to optimize $x_n^p$. The gradient values of $f_2\left({\bm\Psi}^p\right)$ with respect to $x_n^p$ is derived as follows:
\begin{align}
~&\nabla_{x_n^p}{f_2}= \sum_{m = 1}^M \eta \sqrt{1+\alpha_m}\sqrt{p_m}\overline{\beta_{m,n}}D_m-E_m-F_m, \label{gradient_xnp}
\end{align}
where
\begin{subequations}
\begin{align}
D_m&=C_{m,n}\frac{x_m-x_n^p}{\|\mathbf{u}_m-\bm{\psi}_n^p\|^2},\\
E_m&=\sum_{i = m}^M \sum_{n^\prime \neq n}^N \eta p_i\overline{\beta_{m,n}}\beta_{m,n^\prime}\overline{C_{m,n^\prime}}C_{m,n}\frac{x_m-x_n^p}{\|\mathbf{u}_m-\bm{\psi}_n^p\|^2},\\
F_m&=\sum_{i = m}^M 2\eta p_i{\left|\beta_{m,n}\right|}^2\frac{e^{\frac{-2j\pi}{\lambda}\|\mathbf{u}_m-\bm{\psi}_n^p\|^2}(x_n^p-x_m)}{\|\mathbf{u}_m-\bm{\psi}_n^p\|^4}.
 \label{DEF}
\end{align}
\end{subequations}
The proposed GD-based method is summarized in Algorithm \ref{Algorithm_GD}, which is guaranteed to converge to a stationary-point solution \cite{ConvexOp} and $l_\text{min}$ represents the accuracy less than Armijo step.\\

\begin{algorithm}[!t]
    \caption{GD-Based Method for Optimizing $\bm{\Psi}^p$}
    \label{Algorithm_GD}
    Initialize ${\Psi^p}^{(t)}=\left[{\psi_{1}^p}^{(t)},\cdots,{\psi_{N}^p}^{(t)}\right]$ and $t=0$ , step size $l_0$, the minimum step size $l_\text{min}$;\\
    \Repeat{convergence}
    {
        \For{$n=1:N$}
        {
            Calculate $\nabla_{x_n^p}^{(t)}F_2$ and set $l = l_0$\;
            \Repeat{$\left(\ref{limit:x}\right)$ \& $F_2({\mathbf{x}_n^p}^{(t+1)}) > F_2({\mathbf{x}_n^p}^{(t)})$ or $l < l_{\min}$}
            {
                Calculate ${\mathbf{x}_n^p}^{(t+1)} = {\mathbf{x}_n^p}^{(t)} + l \nabla_{x_n^p}^{(t)}F_2$\;
                Set $l = l / 3$\;
            }
            \If{$l < l_{\min}$}
            {
                ${\mathbf{x}_n^p}^{(t+1)} = {\mathbf{x}_n^p}^{(t)}$\;
            }
        }
        Set $t=t+1$\;
    }
\end{algorithm}

\subsubsection{Optimize $\mathbf{P}$}
Finally, by fixing the remaining variables, the optimization problem of $\mathbf{P}$ can be formulated as follows:
\begin{align}
p^\dagger_m = \mathop{\rm argmin}\limits_{{p_m} \leq P_\text{max}}\left({f_3}\left({\textbf P}\right)\triangleq
{p_mB_m-2\Re\left(a_m\right)\sqrt{p_m}}\right),
\end{align}
where $B_m=\!\sum\nolimits_{i=1}^m\! {{\textbf{g}}_m^{\mathsf H}{\bm\beta_i}{\bm\beta_i}^{\mathsf H}{{\textbf{g}}_m}}$, $a_m=\sqrt{1+\alpha_m}{\bm\beta_m}^{\mathsf H}{{\textbf{g}}_m}$.

It's worth noting that maximizing $f_3$ over $\bm P$ is a standard convex optimization problem. By exploiting the first-order optimality conditions, we can obtain the optimized $\bm P$ by setting $\frac{\partial {f_2}}{\partial {p_m}}=0,\space m=1,\dots,M$. After some basic mathematical manipulations, the optimal solution $\bm P^\dagger$ is obtained as follows:
\begin{align} \label{Optimal_P}
p_m^{\dagger}= \min\left\{P_\text{max},\frac{\Re\left(a_m\right)^2}{{B_m}^2}\right\}.
\end{align}

The proposed BCD-based method for solving \eqref{P_2} is summarized in Algorithm \ref{Algorithm_BCD}. We now provide a brief proof for the convergence of our proposed method. For each iteration, we denote a objective value sequence of \eqref{eq:SIC} as $\{{\mathcal{R}}_{\text{sic}}({\textbf P}^{(j)},{\bm{\Psi}^p}^{(j)})\}$, and the objective function of \eqref{P_2} as ${\mathcal{F}_2}\left({\bm \alpha}^{(j)},{\bm \beta}^{(j)},{\textbf P}^{(j)},{\bm{\Psi}^p}^{(j)}\right)$. Then we have
\begin{flalign} \label{Convergence}
&{\mathcal{R}}_{\text{sic}}({\textbf P}^{(j+1)},{\bm{\Psi}^p}^{(j+1)}) \stackrel{\star}{=} {\mathcal{F}_2}\left({\bm \alpha}^{(j+1)},{\bm \beta}^{(j+1)},{\textbf P}^{(j+1)},{\bm{\Psi}^p}^{(j+1)}\right) & \nonumber \\
&\geq {\mathcal{F}_2}\left({\bm \alpha}^{(j)},{\bm \beta}^{(j)},{\textbf P}^{(j)},{\bm{\Psi}^p}^{(j)}\right) \stackrel{\star}{=} {\mathcal{R}}_{\text{sic}}({\textbf P}^{(j)},{\bm{\Psi}^p}^{(j)}).&
\end{flalign}
where ``$\stackrel{\star}{=}$'' holds due to \cite[Section \uppercase\expandafter{\romannumeral3}]{FP}. Note that the sequence $\{{\mathcal{R}}_{\text{sic}}({\textbf P}^{(j)},{\bm{\Psi}^p}^{(j)})\}$ is monotonous for which is upper bounded due to \eqref{limit:p}, thus guaranteeing the convergence of Algorithm \ref{Algorithm_BCD}. Further details can be found in \cite{FP}. The computational complexity of our proposed algorithm can be described based on the dimensions of the problem.

Let $I_{\text{GD}}$ and $I_{\text{BCD}}$ represents the number of the iterations of Algorithm \ref{Algorithm_GD} and \ref{Algorithm_BCD}. The per-iteration computational complexity is composed of updating variables $\{\bm\alpha, \bm\beta, \textbf{P}, \bm{\Psi}^p\}$. With simple proof, the complex of marginal optimizations w.r.t these scale with $\mathcal{O}\left(MN^3\right)$, $\mathcal{O}(MN^3)$, $\mathcal{O}\left(MN\right)$ and $\mathcal{O}\left(I_{\text{GD}}NM^2\log_3(\frac{1}{l_\text{min}})\right)$. Hence, the overall computational complexity of Algorithm \ref{Algorithm_BCD} is approximated as ${\mathcal{O}}\left(I_{\text{BCD}}\left((2MN^3+MN+I_{\text{GD}}NM^2\log_3(\frac{1}{l_\text{min}})\right)\right)$.

\begin{algorithm}[!t]
    \caption{BCD-Based Method For solving $\mathcal{P}_X$}
    \label{Algorithm_BCD}
    Initialize $\left\{{\bm \alpha}^{\left(0\right)},{\bm \beta}^{\left(0\right)},{\textbf P}^{\left(0\right)},{\bm \Psi^p}^{\left(0\right)},\right\}$ and $j=0$\;
    \Repeat{convergence}
    {
        Update $\left\{{\bm \alpha}^{\left(j\right)},{\bm \beta}^{\left(j\right)},{\textbf P}^{\left(j\right)},{\bm \Psi^p}^{\left(j\right)},\right\}$ based on the methods stated in \textbf{1)\space--\space3)}, with the equations of (\ref{Optimal_alpha}),(\ref{Optimal_Beta}),(\ref{Optimal_P})\;
        Set $j=j+1$\;
    }
\end{algorithm}

\subsection{MMSE-NSIC Case}
After solving the uplink sum-rate maximization problem with the MMSE-SIC combining method, we now consider the same optimization goal under the MMSE-nSIC, denoted as ${\mathcal{P}}_{\text{nsic}}$. The objective function for this formulation ${\mathcal{R}}_{\text{nsic}}$ (as defined in \eqref{eq:nSIC}) exhibits structural parallels with the MMSE-SIC version ${\mathcal{R}}_{\text{sic}}$ (defined in \eqref{eq:SIC}). Therefore solving the problem ${\mathcal{P}}_{\text{nsic}}$ share the same procedures and methodologies of solving ${\mathcal{P}}_{\text{sic}}$. Detailed derivations are omitted here for brevity.

\begin{figure}[!t]
\centering
\includegraphics[width=0.4\textwidth]{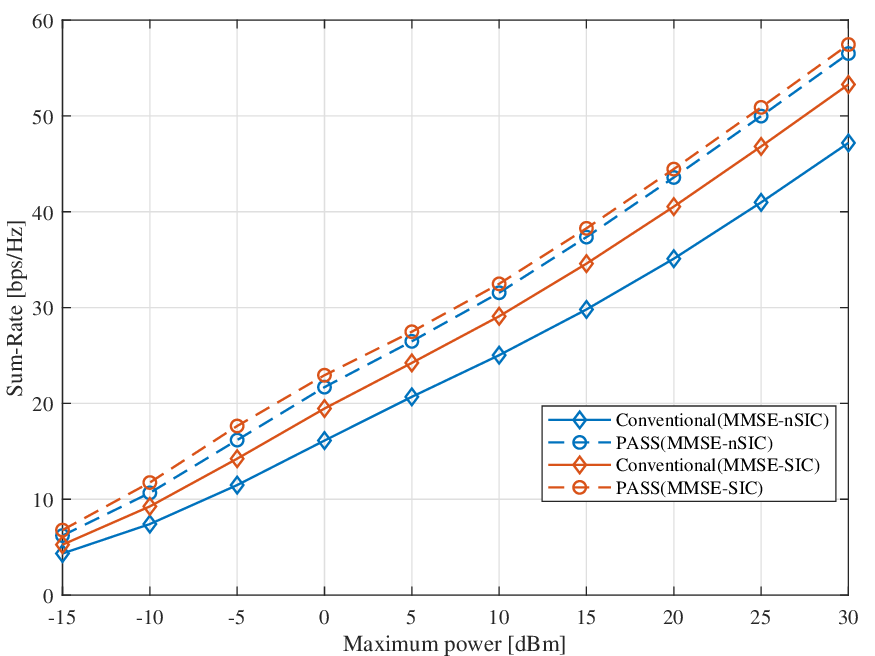}
\caption{The sum-rate versus $P_{\text{max}}$ with $N=4$, $M=4$.}
\label{fig_2}
\end{figure}

\begin{figure}[!t]
\centering
\includegraphics[width=0.4\textwidth]{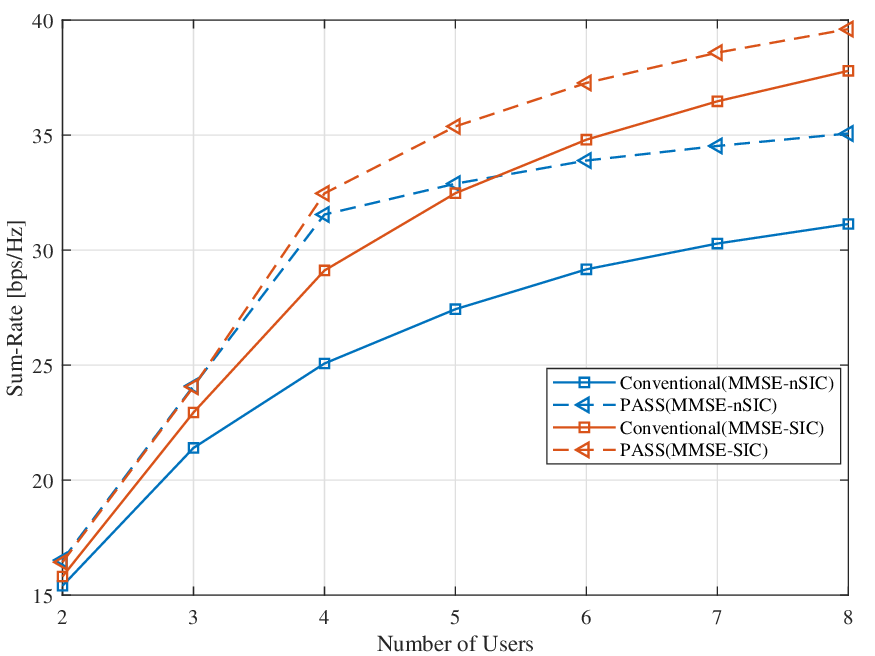}
\caption{The sum-rate versus $M$ with $N=4$, $P_{\text{max}}=10$ dBm.}
\label{fig_3}
\end{figure}

\begin{figure}[!t]
\centering
\includegraphics[width=0.4\textwidth]{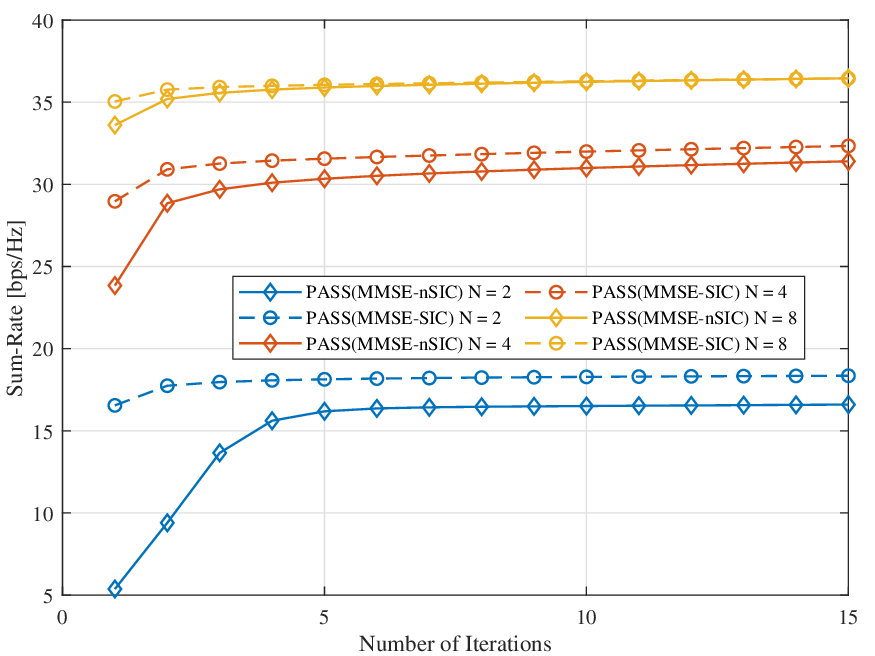}
\caption{Convergence with $M=4$, $P_{\text{max}}=10$ dBm.}
\label{fig_4}
\end{figure}

\section{Numerical results}
This section presents the simulation results of the proposed method with two distinct combining technologies. Specifically, we compare the performance of our proposed method with that of the conventional antenna systems. During the simulation, we set $f_c=28$ GHz, $D_x=15$ m, $D_y=20$ m, $n_{\textrm{eff}}=1.4$\cite{DMMicrowave2021}, $\sigma^2=-90$ dBm, $d=5$ m. The users randomly distribute in the $xy$-plane and PAs are randomly initialized at different waveguides with in $[-D_x,D_x]$. For the conventional-antenna system, we assume that the BS antennas are located at $\tilde{\bm\psi}_n=[0,\tilde{y}_n,d]$, where $\tilde{y}_n =-D_y+\frac{nD_y}{N}$ for $\forall n\in \mathcal{N}=\{1,\dots,N\}$. This can be viewed as a traditional uniform linear array (ULA) placed at the center of area.  All the results are averaged over 10000 independent channel realizations.

Fig. \ref{fig_2} illustrates the impact of users' maximum transmit power on the achievable sum-rate. The results indicate that sum-rate exhibits a monotonically increasing trend with respect to the maximum power allocated to users. Compared to the conventional antenna system with the same combining methods, the proposed PASS has a better performance due to its capability for antenna mobility. The method with SIC is better than nSIC because SIC method considers the proper decoding order to decrease the influence between different users. Notably, within the PASS, SIC method is just a little better than the nSIC method and the performance gap between these two methods ties as the maximum power increases. This demonstrates that through the implementation of the PASS optimization algorithm, the nSIC method with parallel decoding architecture achieves a significantly reduced computational complexity while maintaining the algorithmic performance parity with the SIC approach. This results underscore the critical role played by the PASS optimization algorithm in enabling this performance-complexity trade-off, highlighting its importance in practical communication system design.

Fig. \ref{fig_3} illustrates the observation by showing the sum-rate as a function of the number of users $M$. It's worth noting that in these two methods, the sum-rate grows rapidly as the user count increases when $M\leq N$, while grows slowly when $M>N$. Besides, the performance gap between these two methods widens as the number of users increases. This highlights the importance of using the SIC method when the user count exceeds the number of antennas.

Finally, the convergence of our proposed BCD-based method is illustrated in fig. \ref{fig_4}. As shown, sum-rate converges rapidly for both combiners with different numbers of $N$. As well as the fig. \ref{fig_2} for the PASS, more antennas reduce the difference of performance between the nSIC method and the SIC method. Thus the nSIC method can apply to the scenario of more antennas with low complexity instead of the SIC method.

\section{Conclusion}
Our research has investigated the sum-rate maximization for uplink multiuser MISO PASS. The proposed algorithm jointly optimized PA positions and user power allocation with MMSE-SIC and MMSE-nSIC combining rules. This approach not only achieved better sum-rate performance than conventional systems but also highlighted the flexibility and efficiency of PASS. Through simulations, we have validated the effectiveness of our method and demonstrate substantial gains in sum-rate, especially with MMSE-SIC. Our findings suggested that PASS holds great promise for enhancing the spectral efficiency of uplink multi-user communication systems.

\bibliographystyle{IEEEtran}

\end{document}